\documentclass
[aps,prl,twocolumn,showpacs,preprintnumbers,superscriptaddress,balancelastpage]{revtex4}%
\usepackage[final]{graphicx}
\usepackage{bm}
\usepackage{natbib}
\usepackage{amsfonts}
\usepackage{amsmath}
\usepackage{amssymb}
\usepackage{dcolumn}
\usepackage{latexsym}%
\ifx\pdfoutput\relax\let\pdfoutput=\undefined\fi
\newcount\msipdfoutput
\ifx\pdfoutput\undefined\else
\ifcase\pdfoutput\else
\ifx\paperwidth\undefined\else
\ifdim\paperheight=0pt\relax\else\pdfpageheight\paperheight\fi
\ifdim\paperwidth=0pt\relax\else\pdfpagewidth\paperwidth\fi
\fi\fi\fi
\begin{document}
\preprint{COND\_MAT}
\title{Indirect electric field doping of the CuO$_{2}$ planes in NdBa$_{2}$Cu$_{3}%
$O$_{7}$ cuprate}
\author{M. Salluzzo}
\email{salluzzo@na.infn.it}
\affiliation{CNR-INFM COHERENTIA, Dipartimento di Scienze Fisiche Universit\`{a} di Napoli
"Federico II",Complesso di Monte S. Angelo, Via Cinthia, 80126 Napoli (Italy)}
\author{G. Ghiringhelli}
\affiliation{CNR-INFM COHERENTIA and Dipartimento di Fisica, Politecnico di Milano, piazza
Leonardo da Vinci 32, 20133 Milano, Italy}
\author{J. C. Cezar}
\affiliation{European Synchrotron Radiation Facility, BP220, 38043 Grenoble cedex, France}
\author{N.B. Brookes}
\affiliation{European Synchrotron Radiation Facility, BP220, 38043 Grenoble cedex, France}
\author{G. M. De Luca}
\affiliation{CNR-INFM COHERENTIA, Dipartimento di Scienze Fisiche Universit\`{a} di Napoli
"Federico II",Complesso di Monte S. Angelo, Via Cinthia, 80126 Napoli (Italy)}
\author{F. Fracassi}
\affiliation{CNR-INFM COHERENTIA and Dipartimento di Fisica, Politecnico di Milano, piazza
Leonardo da Vinci 32, 20133 Milano, Italy}
\author{R. Vaglio}
\affiliation{CNR-INFM COHERENTIA, Dipartimento di Scienze Fisiche Universit\`{a} di Napoli
"Federico II",Complesso di Monte S. Angelo, Via Cinthia, 80126 Napoli (Italy)}
\date{\today}

\begin{abstract}
The mechanism of field-effect doping in the \textquotedblleft%
123\textquotedblright\ high critical temperature superconductors \ (HTS) has
been investigated by x-ray absorption spectroscopy in the presence of an
electric field. We demonstrate that holes are created at the CuO chains of the
charge reservoir and that field-effect doping of the CuO$_{2}$ planes occurs
by charge transfer, from the chains to the planes, of a fraction of the
overall induced holes. The electronic properties of the charge reservoir and
of the dielectric/HTS interface determine the electric field doping of the
CuO$_{2}$ planes.

\end{abstract}

\maketitle

The physical phenomena occurring at the interface between materials having
different bulk electronic properties are extremely important in modern
electronics \cite{REVMOD}. The great success of the metal-oxide semiconductor
field effect transistors is due to the high technological control of the
Si/SiO$_{2}$ interface and to the very good understanding of the electric
field effect mechanisms at the interface between a band insulator and a band
semiconductor. Such detailed insight is missing for the interfaces of
materials with strong electron correlation, such as the high critical
temperature superconductors (HTS).

The application of an electric field at the interface between an insulator and
a metal/semiconductor is considered a general method to change reversibly the
electronic properties of thin film without introducing chemical and structural
disorder. In particular, the electric field-effect has been used to shift the
critical temperature and even to induce phase transitions in conventional and
high critical temperature $T_{\text{c}}$ superconductors
\cite{FET1,FET2,FET3,FET4,FET5,FET6}. Experiments conducted on HTS from early
1990s to nowadays have been usually interpreted by supposing that the induced
charges become carriers in the CuO$_{2}$ conducting planes, thus changing the
filling of the Zhang-Rice (ZR) band \cite{ZR}. The universal relationship
between $T_{\text{c}}$ and the carrier density in the CuO$_{2}$ planes of the
HTS (the so-called HTS dome) \cite{DOME}, suggested that an external electric
field could easily lead to controllable changes of their electronic
properties. Although modulations of $T_{\text{c}}$ \cite{FET4,
FETAPLMAN,FETPRLTRI} and of the critical current density \cite{FET5} were
observed in the \textquotedblleft123\textquotedblright\ YBa$_{2}$Cu$_{3}%
$O$_{7}$ and NdBa$_{2}$Cu$_{3}$O$_{7}$ (NdBCO) thin films, the extent of the
electric field effect was found to be below expectations. Superconducting to
insulating transitions were observed only in samples with a doping exactly at
the separation of the two electronic phases \cite{FET3,FET6}, while small
changes on the sample conductivity were observed in Bi$_{2}$Sr$_{1.5}%
$La$_{0.5}$CaCu$_{2}$O$_{8+\delta}$ films having a single insulating CuO$_{2}$
plane doped in a range nominally spanning the whole HTS dome \cite{FETBIECK}.
As matter of fact, a \textbf{direct} doping of the superconducting CuO$_{2}$
planes by the polarization charges created by the electric field has not been
clearly demonstrated yet. \begin{figure}[ptb]
\includegraphics[width=8.0 cm, height=4.3 cm, bb = 0 0 228 125 ]{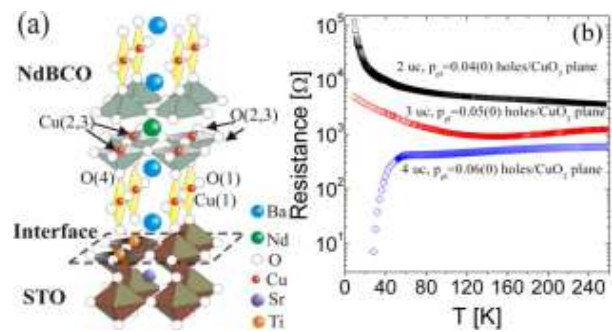}\caption{(Color
online)(a) Sketch of the NdBCO/STO interface with the labeling of the
inequivalent copper and oxygen sites used throughout the text. The interface
plane is shown as a dashed line separating the TiO$_{2}$ plane of the STO
(100) single crystal from the BaO plane of the NdBCO film. b) In situ measured
four point probe channel resistance of the three NdBCO devices: 4 u.c. (green
diamonds) superconducting with critical temperature $T_{\text{c}}=$ 20 K and
carrier density p$_{pl}$ = 0.06(0) holes/CuO$_{2}$ plane; 3 u.c. (red circles)
and 2 u.c. (black circles) insulating with p$_{pl}$ = 0.05(0) and 0.04(0)
holes/CuO$_{2}$ plane, respectively. }%
\label{fig1}%
\end{figure}

Here we present an x-ray absorption spectroscopy (XAS) study of the electric
field doping of the well studied SrTiO$_{3}$/NdBCO interface. The Cu $L_{3}$
edge (2$p_{3/2}\rightarrow3d$ transition) XAS experiment was performed at the
ID08 beam line of the European Synchrotron Radiation Facility. It was designed
in order to allow simultaneous spectroscopic and transport characterization of
the samples under the application of an electric field. Within the wide family
of the HTS, NdBCO belongs to the \textquotedblleft123\textquotedblright%
\ group, where, besides superconducting CuO$_{2}$ planes, an important role is
played by the CuO chains within the \textquotedblleft charge
reservoir\textquotedblright\ (CR). Transmission electron microscopy \cite{TEM}
has demonstrated that the first layer at the SrTiO$_{3}$/NdBCO interface
belongs to the CR, as sketched in Fig.~\ref{fig1}a. This is particularly
relevant since the effective electric field penetration calculated by a Thomas
Fermi model or by any simple electrostatic estimates is of the order of 1 nm
or less, i.e. less than one unit cell (u.c.). In previous studies the role of
the atomic layers directly facing the interface was largely neglected. In this
paper we will show, on the contrary, that they play a decisive role in the
microscopic mechanism of the field effect doping of cuprates.
\begin{figure}[ptb]
\includegraphics[width=7.5 cm, height=5.25 cm, bb = 0 0 214 150 ]{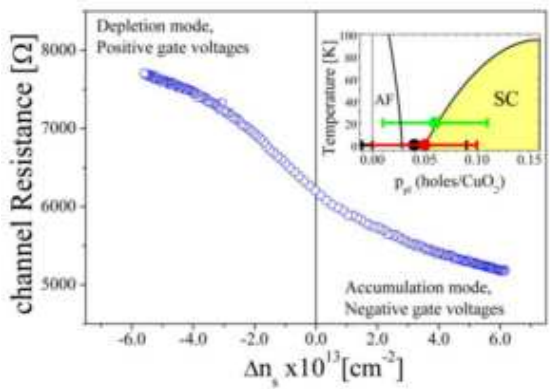}\caption{(Color
online) The channel resistance of the 3 u.c. device, measured as function of
the number of holes per unit area induced by the electric field. In the inset
$T_{\text{c}}$ vs doping (ppl) characteristics of the 4 u.c. (green circle), 3
u.c. (red circle) and 2 u.c. (black circle) overlapped to the HTS dome (SC
stands for superconducting and AF for antiferromagnetic phases). The
horizontal bars around each experimental point are the estimated field effect,
in the assumption that all the induced charge become carriers in the CuO$_{2}$
planes of the first unit cell at the interface.}%
\label{fig2}%
\end{figure}

NdBCO (001) ultra-thin films have been epitaxially grown, by high oxygen
pressure diode sputtering, on TiO$_{2}$ terminated 10x10x 0.5 mm$^{3}$
SrTiO$_{3}$ (100) (STO) surface. The samples are pseudomorphic to the STO
single crystals and have high structural perfection, and very smooth surfaces.
The field effect structures have been completed by deposition on the thin film
of the gold pads for transport measurements, and of the gate electrode
($4\times2$ mm$^{2}$ area) on the back of the substrate, in a geometry
suitable for the XAS experiment. The field has been applied through the
SrTiO$_{3}$ single crystal under ultra high vacuum conditions. The devices
have been characterized in situ during the XAS experiment. A negative
$V_{\text{gate}}$ creates holes at the interface (accumulation mode) and a
positive $V_{\text{gate}}$ reduces the density of holes in the sample
(depletion mode). \ The capacitance of this device is a function of the gate
voltage, a well-known result for non-linear dielectrics materials like STO. At
low gate voltage it is about 2.5 nF and it decreases to 0.4 nF at high
voltages. The charge accumulated at the interface is consequently measured by
an electrometer and is related to the capacitance and to the areal carrier
density $\Delta n_{s}$ through the relation $Q=-e\cdot\Delta n_{s}=\int
_{0}^{V_{gate}}CdV$ . With $V_{\text{gate}}=\pm900$ V, about $\pm
6.0\times10^{13}$ holes/cm$^{2}$ can be subtracted/added at the STO/NdBCO
interface [see also EPAPS file].

We have studied NdBCO samples with a doping above and below the critical value
separating the superconducting and insulating phases (Fig.~\ref{fig1}b)
\cite{SISAL}. An example of the effect of the electric field is shown in
Fig.~\ref{fig2}. The resistivity of the 3 u.c. NdBCO sample at 10 K
non-linearly decreases as function of $\Delta n_{s}$, in agreement with the
induction of carriers in the CuO$_{2}$ planes. It is worth nothing that the
field effect doping is insufficient to induce superconductivity. Similarly, we
could not revert the 4 u.c. sample from superconducting to insulating by
applying a positive gate voltage. These results are in contradiction with
simple model expectations:\ assuming that only one unit cell is doped and that
all the carriers are transferred to the CuO$_{2}$ planes, about p$_{pl}$=0.05
holes/CuO$_{2}$ plane would be added or subtracted. In the 3 u.c. NdBCO the
average hole population per CuO$_{2}$ plane is 0.05(0), consequently the
electric field should bring to zero the conduction carriers for V$_{gate}%
=+900$ V and double them for V$_{gate}=-900$ V. In such a scenario, an
insulating to superconducting transition is expected, as shown in the inset of
Fig.~\ref{fig2}. Even assuming that all the unit cells are doped, a phase
transition should occur anyhow in the samples studied, because the carrier
density would change by $\Delta p_{pl}=\pm$0.015 holes/CuO$_{2}$, i.e., enough
to induce a transition.

\begin{figure}[ptb]
\includegraphics[width=7.5 cm, height=5.2 cm, bb = 0 0 214 150 ]{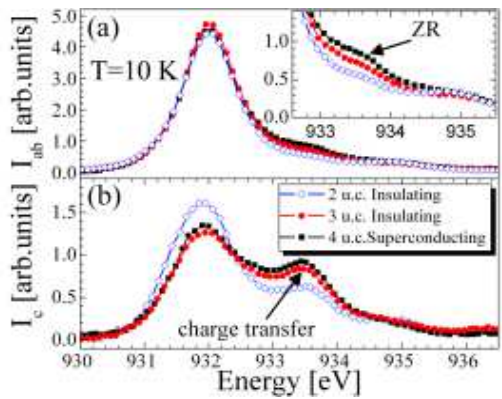}\caption{(Color
online) Cu $L_{3}$ XAS spectra, with $\mathbf{E}\Vert ab$ ($I_{ab}$) and (b)
$\mathbf{E}\Vert c$ ($I_{c}$), measured at low temperature (10 K). The ZR
satellite peak intensity is related to the number of holes per CuO$_{2}$ plane
(see text). This feature decreases continuously when going from 4 u.c. (black
squares) to 2 u.c. (blue diamonds), in agreement with transport data.}%
\label{fig3}%
\end{figure}

To unravel the underlying microscopic mechanism of the electric field effect,
we have used the XAS technique that is sensitive to the orbital occupation and
symmetry at the different copper sites of our samples. We collected the total
electron yield signal that probes a depth of few nanometers, slightly larger
than the NdBCO film thickness. By using the linear polarization of the x-rays,
spectra corresponding to final states lying parallel or perpendicular to the
CuO$_{2}$ planes have been measured ($I_{ab}$ and $I_{c}$ spectra in
Fig.~\ref{fig3}) [see EPAPS file]. 

The Cu $L_{3}$ spectra, measured on the NdBCO films at 10 K, present the
typical features of all cuprates, i.e., a main peak at 932.0 eV and a shoulder
at 933.5 eV (Fig.~\ref{fig3}) \cite{XAS1}. \textquotedblleft%
123\textquotedblright\ cuprates contain two inequivalent copper sites, i.e.,
Cu(1) in the chains and Cu(2,3) in the CuO$_{2}$ planes (see Fig.~\ref{fig1}%
a). As shown in \cite{XAS2} and \cite{XASGHI}, the main peak has a
contribution from the doped Cu(1)$^{3+}$in the chains that form a ligand state
(3$d^{9}\underline{L}$) with the neighboring oxygen ions, which increases if
holes are injected in the chains along the $a-b$ axes. Another contribution
comes from the undoped Cu(2,3)$^{2+}$ in the CuO$_{2}$ planes ($3d^{9}$state).
\ The remaining doped Cu(2,3) sites, occupied by formally trivalent
Cu(2,3)$^{3+}$ ions, give rise to the satellite peak at 933.5 eV. In the
$I_{ab}$ spectra, this feature is due to the Zhang-Rice (ZR) singlet
\cite{ZR}, i.e., to the mobile carriers in the CuO$_{2}$ planes. In the
$I_{c}$ spectra the satellite is due to Cu(2,3)$^{3+}$ sites hybridized with
the apical O(4) oxygen, a signature of the charge transfer from the chains to
the planes. As shown in Fig.~\ref{fig3} and in \cite{XASSAL}, the satellite
peaks decreases with the thickness of the NdBCO films, in agreement with the
transport properties, with the reduction of the carrier density and of the
charge transfer from the chains to the CuO$_{2}$ plane.

\begin{figure}[ptb]
\includegraphics[width=8.0 cm, height=11.0 cm, bb = 0 0 228 313 ] {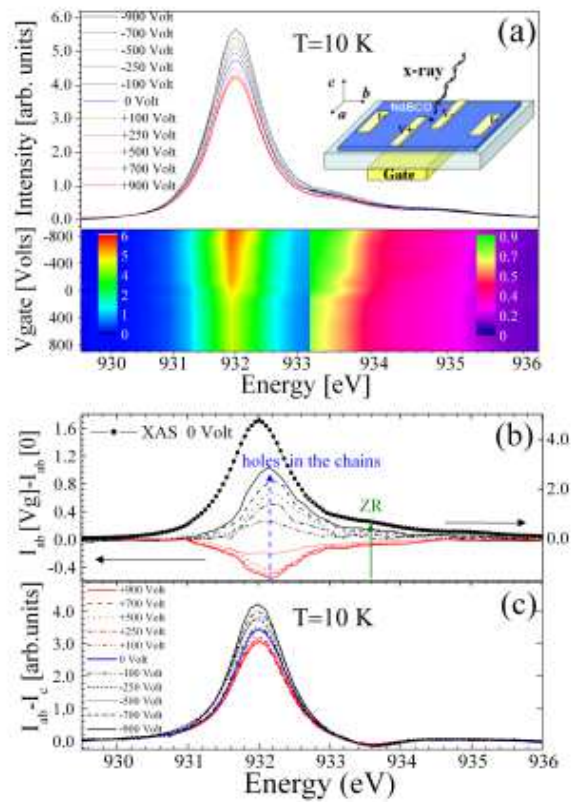}\caption{(Color
on line). (a) Gate voltage dependence of the $I_{ab}$ XAS spectra in the 3
u.c. NdBCO device (the experimental configuration is shown in the inset): in
the upper panel the I$_{ab}$ spectra are shown for negative (black lines) and
positive (red lines) gate voltages. The blue line is acquired at zero field.
In the lower panels the data are shown as color map pictures with different
linear color-scales around the main peak at 932 eV (left panel) and around the
$\underline{c}3d^{10}\underline{L}$ satellite (right panel). (b) Effect of the
field shown by subtracting to each $I_{ab}$ XAS spectra the reference scan
acquired without the field (black closed and line joined circles). (c)
Difference between $I_{ab}$ and $I_{c}$ XAS spectra at different gate
voltages}%
\label{fig4}%
\end{figure}

The effects of the electric field on the $I_{ab}$ spectra at 10 K are shown in
Fig.~\ref{fig4} for the 3 u.c. sample, and similar result were found for the 2
u.c. and 4 u.c. films. Under negative (positive) gate voltages, the satellite
at 933.5 eV, acquires (looses) spectral weight because the injected charges
fill up (empty) the ZR band, i.e., additional carriers appear at the CuO$_{2}$
planes. However, the largest effects are observed on the main peak, which
shows a strong increase (decrease) in intensity and a gradual broadening
(narrowing) on the high-energy side. As shown in Fig. \ref{fig4}b, where the
spectrum at zero field is subtracted from the data, a peak at about 932.2 eV
continuously grows with negative gate voltages. These results are explained by
an enhancement of hole density along the chains by negative fields. The ratio
of the holes induced in the chains (at the Cu(1) sites) and in the CuO$_{2}$
planes (at the Cu(2,3) sites) as estimated from the integrated intensities
around 932.2 eV and 933.5 eV in Fig.~\ref{fig4}b, is $3.5\pm0.5$. It does not
depend on the gate voltage within the experimental error.

The electric field effect is similar in the $ab$ plane and along the $c$ axis.
However, interesting differences can be noted. In Fig.~\ref{fig4}c we show the
XAS ($I_{ab}-I_{c}$) spectra as a function of the field: for negative
(positive) gate voltages, the main peak at 932 eV acquires (looses) relative
in-plane intensity, whereas the satellite anisotropy remains almost
unaffected. This confirms that negative voltages lead to a strong enhancement
of the hole population along the chains (parallel to the $ab$ plane). On the
contrary the filling of the ZR band is accompanied by an equal change of
Cu(2,3)-O(4) ligand states. This is a further confirmation that the doping of
the CuO$_{2}$ planes is rather indirect, i.e., it occurs by a partial charge
transfer from the heavily doped chains via the Cu(2,3)-O(4) bonding.

The XAS results are consistent with the transport data measured in situ and
shown in Fig.~\ref{fig2}. In particular we find a non-linear relationship
between the change in the resistance and the weight of ZR states (proportional
to the integrated intensity of the satellite peak) from V$_{gate}$=+1000 V to
-500 V and a saturation (lower slope) at higher negative voltages. This
resembles the non-linear relationship between the resistivity and induced
carrier density shown in Fig.~\ref{fig2} and can be explained by taking into
account the activated temperature dependence, with a slope depending on the
carrier density, of the resistance in insulating cuprates.

The data are also consistent with previously published results. In
\textquotedblleft123\textquotedblright\ compounds shifts of $T_{\text{c}}$ up
to 10 K where observed in samples not fully superconducting at 4.2 K
\cite{FETAPLMAN,FETPRLTRI}. There the carrier doping at zero field was similar
to ours (about 0.05 holes/CuO$_{2}$ plane) and an increase of about 0.06
holes/CuO$_{2}$ plane should have been expected at maximum field,
corresponding to shift of $T_{\text{c}}$ greater than 30 K, instead of the 10
K observed. In Ref. \onlinecite{FET3} similar differences between the measured
shift in $T_{\text{c}}$ and the values expected from the injected carrier
density are also reported \cite{NOTE2}.

In conclusion, our experimental results give the evidence that an external
electric field injects holes mainly in the chains belonging to the charge
reservoir, whereas only a fraction of them are transferred to the CuO$_{2}$
planes by charge transfer from the charge reservoir to the CuO$_{2}$ planes,
which is the same mechanism governing the doping of the CuO$_{2}$ planes by
chemical substitution in the whole HTS family. This process is specific to the
family of "123" compounds, where the CR is composed by CuO chains. However it
is likely that symilar scheme could operate in the case of other HTS, i.e. the
electric field could modify the valence of the cations in the CR and by in
this way could transfer holes to the CuO$_{2}$ planes.

The similarity between electric field and chemical doping opens new
perspectives for the investigation of the microscopic phase separation
phenomena, proposed to be a key ingredient in the description of
superconducting to insulator transitions of underdoped HTS \cite{SIECK}.
Indeed while the charge induced by field effect is homogeneously distributed
within the gate area at the interface with the thin HTS film, the holes
created by chemical doping are not, especially in the case of insulating or
barely superconducting samples. The apparent similarities between field effect
doping and chemical doping may thus suggest that collective electronic phase
separation is necessary to accommodate doped holes in the parent undoped
compounds. Further studies in this direction are required to confirm this
hypothesis. Finally, our experiment clarifies that, to obtain a shift of
$T_{\text{c}}$ of several degrees or a phase transition, a doping level
substantially higher than usually achieved is necessary. On the other hand, by
having identified the physical mechanisms leading to electric field doping of
the CuO$_{2}$ planes in the HTS, we have shown that an advanced interface
engineering, i.e., the right choice of the atomic layers at the interface and
their chemical and structural control, can eventually open the way to a more
effective field-effect doping of high critical temperature superconductors.

\begin{acknowledgments}
The Authors are grateful to Jean-Marc Triscone and to Antonio Barone for
useful discussion about data interpretation. M. S., G. D. L. and R. V.
acknowledge the support from the EU under the project Nanoxide, contract n. 033191.
\end{acknowledgments}


\begin{references}
\bibitem{NOTE2} In Ref. \onlinecite{FET3} the Hall effect coefficient was used to extract the amount of carriers injected into the CuO$_2$ planes. It is well known that in ``123'' compounds also the chains contribute to the transport, consequently the Hall coefficient depends also on the carrier present in the chains, and, in the case of a field effect experiment, on the carrier injected into the chains. Consequently, it is not surprising to find an agreement between the order of magnitude of the Hall effect coefficient and the estimated field effect injected carrier density.
\end{references}
\end{document}